\documentclass[preprint,amsmath,titlepage,amssymb,english,aps,showkeys]{revtex4}

\usepackage{xcolor}
\usepackage{amssymb,amsmath}
\usepackage{graphicx}
\usepackage{hyperref}
\usepackage{xcolor}
\usepackage[utf8]{inputenc}
\usepackage[T1]{fontenc}
\usepackage{amsfonts}

\begin{document}

\title{A Five-dimensional Kaluza-Klein Approach to Unimodular Gravity }

\author{J{\'u}lio C. Fabris}
\email{julio.fabris@cosmo-ufes.org}
\affiliation{Núcleo Cosmo-ufes\&Departamento de F{\'i}sica, CCE, Universidade Federal do Espírito Santo, Vitória, ES, Brazil}

\author{ Richard Kerner}
\email{richard.kerner@sorbonne-universite.fr}
\affiliation{Laboratoire de Physique Th\'eorique de la Mati\`ere Condens\'ee, Sorbonne-Universit\'e, Boite 121, 
4 Place Jussieu, 75005, Paris,  France}

\date{\today}

\begin{abstract}

In this article we present a possibility of imposing the unimodular condition within the $5$-dimensional Kaluza-Klein theory including the
scalar field. Unimodular gravity became an object of increasing interest in the late $80$-ties (see, e.g. \cite{Henneaux}, \cite{Dragon}, \cite{Unruh});
and was recently used in primordial Universe modeling with cosmological constant, in the context of the Brans-Dicke
gravity including scalar field (\cite{Almeida}). A generalization of the unimodularity principle to the $5$-dimensional Kaluza-Klein model
was discussed in our recent paper (see \cite{FabrisKerner2024}, \cite{FabrisKerner2025}), in which variational principle is formulated in $5$ dimensions first, 
and dimensional reduction is applied to the resulting set of equations. A cosmological model based on these equations was then presented and discussed.
Here we present further developments of this approach, focussing our attention at perturbative aspects and stability of solutions.

\end{abstract}

\keywords{Kaluza-Klein theories, Unimodular gravity, Cosmology, Perturbation analysis}

\maketitle

\section{Introduction}

Soon after the mathematical foundations of the General Relativity theory were set forth by Einstein in $1915$ (\cite{Einstein1915}), 
the search for solutions began. It is worth mentioning that the first successful applications of new theory of gravitation, namely 
the perihelion advance and the bending of light in strong gravitational field, were using approximations, and not exact solutions, 
inexistent at the time. Both effects are so small, that exact solutions would not have brought any practical improvement, the effects 
of the next order being below the perception of detecting devices. The first exact solution of Einstein's equations {\it in vacuo}:
\begin{equation}
G_{\mu \nu} = R_{\mu \nu} - \frac{1}{2} g_{\mu \nu} R = 0.
\label{Einsteineqs}
\end{equation}
was found by Karl Schwarzschild in $1916$ (\cite{Schwarzschild1916}). In $1917$ Einstein (\cite{Einstein1917}) modified his equations 
introducing an extra term proportional to the metric tensor $g_{\mu \nu}$, in order to make a static cosmological solution possible:
\begin{equation}
R_{\mu \nu} - \frac{1}{2} g_{\mu \nu} R - \Lambda g_{\mu \nu} = - \frac{8 \pi G}{c^4} \; T_{\mu \nu}
\label{Einstein1917}
\end{equation}
where $T_{\mu \nu}$ is the covariantly conserved energy-momentum tensor, $\nabla^{\mu} T_{\mu \nu} = 0$, where the covariant derivation 
is supposed to be performed with respect to the Christoffel connection derived from the pseudo-Riemannian metric $g_{\mu \nu}$. 
The metric tensor $g_{\mu \nu}$ is also covariantly constant, i.e. we have by definition $\nabla_{\lambda} g_{\mu \nu} = 0$, and obviously, 
also the covariant $4D$ divergence of $g_{\mu \nu}$ also vanishes. Consequently, the cosmological term $\Lambda \; g_{\mu \nu}$ can be added 
to the energy-momentum tensor without changing its conservative character:
\begin{equation}
\nabla^{\mu} ( T_{\mu \nu} + \Lambda g_{\mu \nu} ) = 0 \; \; \; {\rm if} \; \; \;  \nabla^{\mu}  T_{\mu \nu}  = 0.
\label{Tconserve}
\end{equation} 

In $1919$ Einstein proposed a constrained version of General Relativity by imposing the unimodularity condition on the metric,
\cite{ein} in order to fix a coordinate system. The metrics considered as possible solutions should have their determinant
equal to $1$, coinciding with the Minkowski relation in cartesian coordinates, 
that is, $\sqrt{-g} = 1$.  At that time the Friedmann solution was still unknown, and the unimodularity
condition was an elegant way to justify the presence of the cosmological term, appearing as a Lagrange multiplier. 
After Friedmann's solution describing a time-dependent Universe was found, Einstein postponed the cosmological constant term
judging it as unnecessary, and lost interest in unimodular gravity as well.

However, unimodular gravity became an object of increasing interest in the late $80$-ties (see, e.g. \cite{Henneaux}, \cite{Dragon}, \cite{Unruh});
and was recently used in primordial Universe modeling with cosmological constant, in the context of the Brans-Dicke
gravity including scalar field (\cite{Almeida}).  

It is difficult to say where the idea of extra spatial dimensions could come from, but in
the early $20^{\rm th}$ century Theodor Kaluza ($1885-1954$) {\cite{Kaluza}} proposed a unified theory of gravity and electromagnetism
 based on the Einsteinian General Relativity extended to five dimensions. By adding
an extra spatial coordinate $x^5$ and assuming that Riemannian geometry applies to the enlarged manifold,
Kaluza noticed that the $15$ independent components of metric tensor in $5$ dimensions can acomodate the $10$ components
of a $4$-dimensional subspace which can be identified with the metric tensor of General Relativity, but also the extra
four components of the $4$-potential of Maxwell's electromagnetism, if they are identified with mixed components 
${\tilde{g}}_{5 \mu} = {\tilde{g}}_{\mu 5}$, where $\mu, \nu = 0,1,2,3$:
\begin{equation}
{\tilde{g}}_{AB} = \begin{pmatrix} g_{\mu \nu} & A_{\mu} \cr A_{\nu} & 1 \end{pmatrix}.
\label{gKK}
\end{equation} 
( In what follows, we shall denote by tilded symbols all geometrical objects
defined in the $5$-dimensional Kaluza-Klein space, the same symbols relative to the $4$-dimensional space-time being given as usual,
by non-tilded letters.)

The standard variational principle applied to the Einstein-Hilbert lagrangian in $5$ dimensions lead to $15$ partial differential equations;
therefore there is a risk that such a system would be over-determined, given that we have only $14$ independent fields, $g_{\mu \nu}$
and $A_{\mu}$. 

However, by a happy coincidence, even with this incomplete version, the system was not over-determined due to the fact
that out of the $15$ Einstein equations in vacuo corresponding to the components of symmetric Ricci and metric tensors:
$(\mu \nu), \; \; (\mu 5 )$ and $(5 5)$ the last one ${\tilde{R}}_{55} - \frac{1}{2} {\tilde{g}}_{55} {\tilde{R}}$ reduces to tautology $0 = 0$, 
leaving exactly $14$ equations, which are easily recognized as the usual $4$-dimensional Einstein's equations with electromagnetic energy-momentum
tensor as a source, along with Maxwell's equations coupled with gravitational field through covariant derivatives: the $15$ equations
\begin{equation}
{\tilde{R}}_{AB} - \frac{1}{2} {\tilde{g}}_{AB} {\tilde{R}} = 0, \; \; (A, B,..) = (\mu, 5), \; \; 
{\tilde{R}} = R - \frac{1}{4} F_{\lambda \rho} F^{\lambda \rho}.
\label{fiveeqs}
\end{equation}
with an extra assumption that the fields $g_{\mu \nu}$ and $A_{\mu}$ depend exclusively on space-time variables $x^{\mu}$ and not on $x^5$, 
when explicited give rise to the following system of coupled equations:
\begin{equation}
R_{\mu \nu} - \frac{1}{2}g_{\mu \nu} R = g^{\lambda \rho} F_{\mu \lambda} F_{\nu \rho} - \frac{1}{4} g_{\mu \nu} F_{\lambda \rho} F^{\lambda \rho},
\label{KKmunu}
\end{equation}
\begin{equation}
g^{\mu \nu} \nabla_{\mu} F_{\nu \lambda} = 0, 
\label{Maxwellcov}
\end{equation}
the last $55$ component reducing to $0 = 0$. This circumstance is often called ``the Kaluza-Klein miracle''.

But it is not the only one. There is another happy coincidence, namely, the fact that the determinant of the $5$-dimensional Kaluza-Klein
metric does not depend on the electromagnetic fields $A_{\mu}$, but is a product of the determinant of the $4$-dimensional metric of the
space-time multiplied by a function of the scalar field. This circumstance encourages the idea of introducing the unimodular condition in the
$5$-dimensional Kaluza-Klein space. 

In its first version proposed by Th. Kaluza, the fifth dimension was just an extra space coordinate, the
entire space being isomorphic with $M_4 \times R^1 \sim [ct, x, y, z, x^5]  \sim M_5$, a five-dimensional Minkowski space. 

A few years later, after the advent of Quantum Mechanics, Oskar Klein (\cite{Klein}) proposed to consider a compact fifth dimension, a circle with 
a very small radius. 
\vskip 0.5cm
\begin{figure}[hbt]
\centering
\includegraphics[width=4.4cm, height=2.5cm]{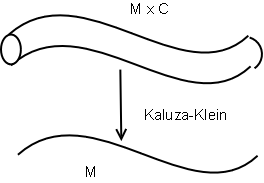} 
\label{fig:KKscheme}
{\caption{\small { The five-dimensional Kaluza-Klein space with a compact $5^{\rm th}$ dimension. The local coordinates are 
$x^A = (x^{\mu}, x^5); \; \; A = 1,2,..5, \; \; \mu, \nu,.. = (0, i) = 0,1,2,3, $
which, under the projection $\pi$, reduce to points in the Minkowski space-time:
$\pi (x^A) = \pi (x^{\mu}, x^5) = (x^\mu) \in M_4. $ }}}
\end{figure}

The dependence on the fifth dimension of functions defined on the  ``compactified'' space must be therefore periodic,
with a Fourier-like decomposition: 
\begin{equation}
f(x^{\mu}, x^5) = {\displaystyle{\sum_{k=0}^{\infty} }} a_k (x^{\mu}) e^{ik m x^5}.
\label{eik}
\end{equation}
 with dim(m) =cm$^{-1}$.

Consequently, the eigenvalues of the fifth component of quantum momentum operator,  $p_5 = - i \hbar \partial_5$ 
are integer multiples of mass $m$, or perhaps the electric charge if one chooses an appropriate physical dimensional parametrisation.
Klein's aim was to explain the discrete values of electric charge of elementary particles known at that time, electrons and protons. 

A unimodular version of the Kaluza-Klein model was considered recently in our last papers (see \cite{FabrisKerner2024}, \cite{FabrisKerner2025}). 
We considered cosmological solutions
of the modified model, incorporating the scalar fields realized as the $g_{55}$ component of the $5$-dimensional metric tensor. Some aspects
of  small perturbations were discussed, too. Here we present a more detailed analysis of cosmological solutions
and their stability with respect to small perturbations.

A couple of decades later the Kaluza-Klein theory was improved by P. Jordan \cite{Jordan} and Y. Thiry \cite{Thiry}
by inclusion of scalar field and its impact on generalized solutions. 

 Although the five-dimensional Kaluza-Klein theory is not viable, one of the reasons being the absence 
of chiral spinors in five space-time dimensions, it is present in some way or another in a more realistic theory 
as a restriction to the $U(1)$-subgroup of the full structural group. As such, it describes fairly well the equations 
governing the electromagnetic sector of any unified theory when other interactions (strong and electroweak ones) can be neglected.
 
In what follows, we shall formulate the variational principle in $5$ dimensions first, and apply dimensional reduction to
the resulting set of equations. The properties of the resulting $4D$-model based on these equations will then be presented and discussed.

\section{Unimodular gravity}

Unimodular gravity was first proposed by Einstein in 1919 \cite{ein} in order to fix a coordinate system
by imposing a condition on the determinant of the metric, coinciding with the Minkowski relation in cartesian coordinates, 
that is, $\sqrt{-g} = 1$. This proposal has evolved more recently and the unimodular gravity was viewed as another approach 
to the cosmological constant problem and to quantization of gravity, see for example \cite{Unruh,wein,brito,bran,u1}. 
In generalizing the unimodular gravity to the Kaluza-Klein, we follow the lines sketched in \cite{bran} (see also \cite{u2}).

Prior to describing the implementation of the unimodular constraint in the Kaluza-Klein theory, 
let us revise briefly its possible implementations in the context of General Relativity theory through the addition of the constraint
in the Einstein-Hilbert action using the Lagrange multiplier technique. 
Let us consider the action
\begin{eqnarray}
{\cal S} = \int d^4x\biggr\{\sqrt{-g}R - \chi(\sqrt{-g} - \xi)\biggl\} + \int d^4 x \sqrt{-g}{\cal L}_m.
\end{eqnarray}
In this expression, $\chi$ is a Lagrange multiplier, and $\xi$ is an external field introduced in order to give flexibility in the choice of the coordinate system. 
As a special case, $\xi$ may be considered a pure number, which implies the choice of a specific coordinate system.  

The variation with respect to the metric leads to:
\begin{eqnarray}
\label{erg1}
R_{\mu\nu} - \frac{1}{2}g_{\mu\nu}R + \frac{\chi}{2}g_{\mu\nu} = 8\pi GT_{\mu\nu}.
\end{eqnarray}
The variation with respect to $\chi$ leads to the unimodular constraint: 
\begin{eqnarray}
\label{vin-rg-1}
\xi = \sqrt{-g}.
\end{eqnarray} 

The trace of (\ref{erg1}) implies:
\begin{eqnarray}
\chi = \frac{R}{2} + 8\pi G\frac{T}{2}.
\end{eqnarray}
Inserting this result in (\ref{erg1}), we obtain the unimodular field equations:
 \begin{eqnarray}
 \label{erg2}
 R_{\mu\nu} - \frac{1}{4}g_{\mu\nu}R = 8\pi G\biggr(T_{\mu\nu} - \frac{1}{4}g_{\mu\nu}T\biggl).
 \end{eqnarray}
The theory is now invariant by a restricted class of diffeomorphisms, called transverse diffeomorphisms. 
 
The Bianchi identities imply that, in general, the usual energy-momentum tensor conservation must be generalized as follows:
 \begin{eqnarray}
 \label{brg1}
 \frac{R^{;\nu}}{4} = 8\pi G\biggr({T^{\mu\nu}}_{;\mu} - \frac{1}{4}T^{;\nu}\biggl).
 \end{eqnarray} 

If the usual energy-momentum tensor conservation is imposed (which in the unimodular context, would be in opposition with basic General Relativity theory), 
 \begin{eqnarray}
 \label{cons-rg-1}
 {T^{\mu\nu}}_{;\mu} = 0,
 \end{eqnarray}
 the equation (\ref{brg1}) becomes
 \begin{eqnarray}
 \label{brg2}
 \frac{R^{;\nu}}{4} = -2 \pi GT^{;\nu}.
 \end{eqnarray} 
The above equation (\ref{brg2}) may be integrated leading to, 
\begin{eqnarray}
\label{ic1}
R = - 2\pi GT + 4\Lambda,
\end{eqnarray}
where $\Lambda$ is an integration constant which may be interpreted as the cosmological constant. 
Inserting the relation (\ref{ic1}) in (\ref{erg1}), the resulting equations are: 
\begin{eqnarray}
 \label{erg3}
 R_{\mu\nu} - \frac{1}{2}g_{\mu\nu}R = 8\pi GT_{\mu\nu} + g_{\mu\nu}\Lambda.
 \end{eqnarray}
We see that the equations of standard General Relativity are recovered, but with a cosmological constant which was absent in the original structure, but
emerges naturally as a constant of integration.   

\section{Full model including the scalar field}

Let us remind the full version of the Kaluza-Klein model, which englobes gravitational field given by the $4$-dimensional metric 
$g_{\mu \nu} (x)$, the electromagnetic field given by its $4$-potential $A_{\mu} (x)$ and the scalar field $\Phi (x)$, 
This ansatz corresponds to the $15$ degrees of freedom present in the $5$-dimensional Kaluza-Klein symmetric
metric tensor ${\hat{g}}_{AB}, \; \; 1,2, ... 5$.

The component $g_{55}$ of the five-dimensional metric tensor should be strictly negative in order to keep the fifth dimension spatial; 
this is why we shall give it the form $g_{55} = - \Phi^2 (x)$

Starting from this assumption, three particular situations can be studied separately now. We can consider a case with scalar field only being present, 
without the electromagnetic one.
This will lead to a variant of the tensor-scalar theory of gravitation, similar to the one proposed by Brans and Dicke (\cite{Brans}).

Another choice is the classical Kaluza-Klein model unifying gravitation and electromagnetism, but without scalar field. 
This choice amounts to suppressing one degree of freedom out of $15$, leaving only $14$ ones, 
the $4$-dimensional space-time metric $g_{\mu \nu}$
and the $4$-vector potential encoded in the components ${\hat{g}}_{\mu 5} = {\hat{g}}_{5 \mu}$ 
of the $5$-dimensional metric. 

Finally, we may consider the electromagnetic and scalar fields interacting in a flat Minkowskian space-time, 
in the absence of gravitation field considered as being negligible.

The five-dimensional metric with scalar field $\Phi(x)$ as the single degree of freedom remains diagonal:
\begin{equation}
g_{AB} = {\rm diag} \left(+1, -1, -1, -1, -\Phi^2(x) \right).
\label{gABdiag}
\end{equation} 
In principle, the notation $\Phi(x)$ can mean the dependence of the scalar field $\Phi$ not only on the space-time coordinates
$ (x^0=ct, x^1, x^2, x^3) $ but also on the fifth coordinate $x^5$, so that in principle we may have 
not only $\partial_{\mu} \Phi \neq 0$, but also $\partial_5 \Phi \neq 0$.

However, supposing that the fifth dimension is the structural group $U(1)$, i.e. is homeomorphic to a circle $S^1$, the dependence
of $\Phi$ on $x^5$ can be only a periodic one:
\begin{equation}
\Phi ( x^{\mu}, x^5 ) = \cos (\tilde n \; e \; x^5 + \delta ) \cdot \Phi (x^{\mu}), \; \; {\rm so \; that} \; \; \partial_{5}^2 \Phi = - \tilde n^2 e^2 \Phi,
\label{partial55}
\end{equation}  
where $\tilde n$ correspond to the normal mode. The full set of general formulas for metrics, connections and curvature in $5$ dimensions, 
with all the $15$ degrees of freedom present, in an appropriate non-holonomic frame, is given in the Appendix at the end of the article.

What we need here is the expression for the $5$-dimensional scalar curvature $\hat{R}$ expressed in terms of the $4$-dimensional metric, 
the electromagnetic potential and the scalar field. We shall consider $\hat{R}$ as integrand in the five-dimensional variational principle.
As shown in the Appendix, this quantity can be expressed by means of the familiar four-dimensional geometric objects:
\begin{equation}
{\hat{R}} = {\overset{4}{R}} - \frac{1}{4} \Phi^2 \; F_{\mu \nu} F^{\mu \nu} - \frac{2}{3 \Phi^2} g^{\mu \nu} \partial_{\mu} \Phi \partial_{\nu} \Phi.
\label{Rexpl}
\end{equation}
Considered as the integrand of a $5$-dimensional variational principle, this Lagrangian density will lead to the following Einstein's equations
when varying with respect to the metric only:
\begin{equation} 
{\hat{R}}_{AB} - \frac{1}{2} {\hat{g}}_{AB} {\hat{R}} = 8 \pi G \left[ T_{AB}^{(\Phi)} + \frac{k^2}{16 \pi G} T_{AB}^{(F)} \right]
\end{equation}
where formally
\begin{equation}
T_{AB}^{(\Phi)} = \partial_A \Phi \partial_B \Phi - \frac{1}{2} {\hat{g}}_{AB} ({\hat{g}}^{CD} \partial_C \Phi \partial_D \Phi ),
\label{TPhi}
\end{equation}
and 
\begin{equation}
T_{AB}^{(F)} = F_{AC} F^C_{\; \; B} - \frac{1}{4} {\hat{g}}_{AB} (F_{CD} F^{CD} ),
\label{TPeF}
\end{equation}
which in the case of the $\tilde n-th$ mode, i. e.,  the dependence $\Phi$ on $x^5$ in a periodic way, only to the space-time components
different from zero:
\begin{equation}
T_{\mu \nu}^{(\Phi)} = \partial_{\mu} \Phi \partial_{\nu} \Phi - \frac{1}{2} {\hat{g}}_{\mu \nu} \left[ {\hat{g}}^{\lambda \rho} 
 \partial_{\lambda} \Phi \partial_{\rho} \Phi - n^2 e^2 \Phi^2 \right] ,
\label{TPhimunu}
\end{equation}
(where we neglected the mixed terms with $F_{\mu \nu}$ ) and 
\begin{equation}
T_{\mu \nu}^{(F)} = F_{\mu \lambda} F^{\lambda}_{\; \; \nu} - \frac{1}{4} {\hat{g}}_{\mu \nu} (F_{\lambda \rho} F^{\lambda \rho} ).
\label{TPeFmunu}
\end{equation}

The variation with respect to the scalar field $\Phi$ and the $4$-vector potential $A_{\mu}$ lead to the following equations of motion:

\begin{equation}
\frac{1}{\Phi} \partial_{\mu} \left[ \Phi \; F^{\mu \nu} \right] = 0,
\label{MotionF}
\end{equation}
and 
\begin{equation}
(\Box \Phi + \tilde n^2 e^2 ) \Phi = 0.
\label{boxPhi} 
\end{equation}
where the term $\tilde n^2 e^2$ comes from the second derivative of $\Phi$ with respect to the circular coordinate $x^5$ 
and plays the role of a mass term for the Klein-Gordon scalar field equation.

\section{ Kaluza-Klein unimodular gravity }


Our investigation of unimodular gravity in the context of the $5$-dimensional Kaluza-Klein theory should be
preceded by the following considerations. We are confronted with an alternative choice of the way we apply the unimodularity
condition. We can choose to impose it to the variational principle in $5$ dimensions, find the solutions and then proceed
to dimensional reduction, or perform dimensional reduction first, expressing everything in terms of $4$-dimensonal Riemannian
metric and electromagnetic and scalar fields, and then proceed to variational principle in $4$ dimensions with unimodular
condition imposed on the $4$-dimensional metric. 


In order to explore all the degrees of freedom present in the Kaluza-Klein setting, we choose to apply the unimodular condition
directly to the $5$-dimensional variational principle. Our starting point is the following Lagrangian density:
\begin{eqnarray}
{\cal L} = \sqrt{\tilde{g}}{\tilde{R}} + \lambda(\sqrt{\tilde{g}} - \xi) + \sqrt{\tilde{g}} {\cal L}_m.
\end{eqnarray}
The tildes indicate the five dimensional quantities; $\lambda$ is the Lagrangian multiplier fixing the unimodular condition.

Using the variational principle:
\begin{eqnarray}
\label{eq1}
\tilde R_{AB} - \frac{1}{2}\tilde g_{AB}\tilde R - \frac{\lambda}{2}g_{AB} &=& 8\pi G T_{AB}, \\
\label{eq2}
\sqrt{\tilde g} &=& \xi.
\end{eqnarray}
The external function (not subject to the variational principle) $\xi$ allows to fix the determinant of the volume 
in a given coordinate system.

Taking the trace of (\ref{eq1}), it results in the following relation:
\begin{eqnarray}
\lambda = \frac{2 - D}{D}\tilde R - \frac{16\pi G}{D} T,
\end{eqnarray}
where $D = 4 + n$, $n$ being the number of extra-dimension.

Inserting this relation again into (\ref{eq1}), we obtain:
\begin{eqnarray}
\label{eq-u1}
\tilde R_{AB} - \frac{1}{D}\tilde g_{AB}\tilde R &=& 8\pi G \biggr(\tilde T_{AB} - \frac{1}{D}g_{AB}\tilde T\biggl), \\
\sqrt{\tilde g} &=& \xi.
\end{eqnarray}

Using the Bianchi identities in (\ref{eq-u1}), we obtain,
\begin{eqnarray}
\frac{D - 2}{2D}\tilde R^{;A} = 8\pi G\biggr\{{\tilde T^{AB}}_{;B} - \frac{\tilde T^{;A}}{D}\biggl\}.
\end{eqnarray}
If the the conservation of the energy-momentum tensor is imposed, ${\tilde T^{AB}}_{;B} = 0$, it is possible to write,
\begin{eqnarray}
\frac{D - 2}{2D}\tilde R + 8\pi G\frac{\tilde T}{D} = - \tilde\Lambda,
\end{eqnarray}
where $\tilde\Lambda$ is an integration constant which can be identified with the multidimensional cosmological constant.
The field equations become then,
\begin{eqnarray}
\label{eq-u2}
\tilde R_{AB} - \frac{1}{2}\tilde g_{AB}\tilde R = 8\pi G \tilde T_{AB} + g_{AB}\tilde\Lambda.
\end{eqnarray}
The imposition of the conservation of the energy-momentum tensor is, however, an extra hypothesis.
In order to preserve the unimodular condition (\ref{eq2}) under an infinitesimal coordinate transformation, 
\begin{eqnarray}
x^A \quad \rightarrow \quad x^A + \epsilon^A,
\end{eqnarray}
the transversal condition
\begin{eqnarray}
\epsilon^A_{;A} = 0.
\end{eqnarray}
This restriction generalizes, in $D$ dimension, the corresponding condition in four dimension: the unimodular theory in higher dimensions 
is still invariant by the transversal class of diffeomorphism (for a discussion on transversal diffeomorphism, see for example \cite{td}).

\section{Reduction to four dimensions: the cosmological setup}

Initially, the Kaluza-Klein model was intended to incorporate Maxwellian electromagnetism into Einstein's General Relativity. The first cosmological applications
can be traced down to the Brans-Dicke scalar-tensor theory which turned out to be isomorphic with a variant of Kaluza-Klein theory including a scalar field
as the $(55)$ component of metric tensor. The explicit cosmological ansatz in $4+n$ dimensions, along with the generalized Einstein's equations, are given
in Appendix II.

Let us proceed by constructing a four dimensional theory out of the multidimensional theory. We change slightly the notation used in Appendix II.
We write the multidimensional metric, with dimension $D = 4 + n$, as,
\begin{eqnarray}
ds^2 = g_{\mu\nu}dx^\mu dx^\nu - \Phi^2\delta_{ab}dx^a dx^b. 
\end{eqnarray}
The non-null components of the Christoffel symbols
\begin{eqnarray}
\tilde\Gamma^C_{AB} = \frac{\tilde g^{CD}}{2}\biggr(\partial_A \tilde g_{DB} + \partial_B\tilde g_{DA} - \partial_D\tilde g_{AB}\biggl),
\end{eqnarray}
are:
\begin{eqnarray}
\tilde\Gamma^\rho_{\mu\nu} &=& \Gamma^\rho_{\mu\nu},\\
\tilde\Gamma^{\rho}_{ab} &=& \Phi\Phi^{;\rho}\delta_{ab},\\
\tilde\Gamma^a_{\rho b} &=& \frac{\Phi_{;\rho}}{\Phi}\delta_{ab}.
\end{eqnarray}

The non-null components of the Ricci tensor are:
\begin{eqnarray}
\tilde R_{\mu\nu} &=& R_{\mu\nu} - n\frac{\Phi_{;\mu;\nu}}{\Phi}, \\
\tilde R_{ab} &=& \biggr\{\Phi\Box\Phi + (n - 1)\Phi_{;\rho}\Phi^{;\rho}\biggl\}\delta_{ab}.
\end{eqnarray}
The Ricci scalar is:
\begin{eqnarray}
\tilde R = R - 2n\frac{\Box\Phi}{\Phi} - n(n - 1)\frac{\Phi_{;\rho}\Phi^{;\rho}}{\Phi^2}.
\end{eqnarray}

The components of the gravitational tensor
\begin{eqnarray}
\tilde G_{AB} = \tilde R_{AB} - \frac{1}{2}\tilde g_{AB}\tilde R,
\end{eqnarray}
are:
\begin{eqnarray}
\tilde G_{\mu\nu} &=& G_{\mu\nu} - n \biggr\{\frac{\Phi_{;\mu;\nu}}{\Phi} - g_{\mu\nu}\frac{\Box\Phi}{\Phi}\biggl\} 
+ \frac{n(n - 1)}{2}g_{\mu\nu}\frac{\Phi_{;\rho}\Phi^{;\rho}}{\Phi^2},\\
\tilde G_{ab} &=& \frac{\Phi^2}{2}\biggr\{R - 2(n - 1)\frac{\Box\Phi}{\Phi} 
 - (n - 1)(n - 2)\frac{\Phi_{;\rho}\Phi^{;\rho}}{\Phi^2}\biggl\}\delta_{ab}.
\end{eqnarray}

The components of the unimodular gravitational tensor
\begin{eqnarray}
\tilde E_{AB} = \tilde R_{AB} - \frac{1}{D}\tilde g_{AB}\tilde R,
\end{eqnarray}
are,
\begin{eqnarray}
\tilde E_{\mu\nu} &=& R_{\mu\nu} - \frac{1}{n+ 4}g_{\mu\nu}R - n\frac{\Phi_{;\mu;\nu}}{\Phi} + \frac{1}{n + 4}g_{\mu\nu}\biggr\{2n\frac{\Box\Phi}{\Phi}
 + n(n - 1)\frac{\Phi_{;\rho}\Phi^{;\rho}}{\Phi^2}\biggl\},\\
\tilde E_{ab} &=&  \frac{\Phi^2}{n + 4}\biggr\{R - (n - 4)\frac{\Box\Phi}{\Phi} + 4(n - 1)\frac{\Phi_{;\rho}\Phi^{;\rho}}{\Phi^2}\biggl\}\delta_{ab}.
\end{eqnarray}

\subsection{Introducing matter}

The energy-momentum tensor for a fluid in $n$ dimensions is given by,
\begin{eqnarray}
T^{AB} = (\rho + p)u^A u^B - p g^{AB}.
\end{eqnarray}
We define,
\begin{eqnarray}
\tilde\tau_{AB} = T_{AB} - \frac{1}{D}g_{AB}T.
\end{eqnarray}
We use co-moving coordinates such that,
\begin{eqnarray}
u^A = (1,\vec 0).
\end{eqnarray}

We will distinguish the pressure in the external space, denoted by $p_e$, and the pressure in the internal space, denoted by $p_i$.
A direct computation leads to,
\begin{eqnarray}
\tau_{00} &=& \frac{1}{n + 4}\biggr\{(n + 3)\rho + 3 p_e + n p_i\biggl\},\\
\tau_{ij} &=& \frac{a^2\delta_{ij}}{n + 4}\biggr\{\rho + (1 + n)p_e - n p_i\biggl\},\\
\tau_{ab} &=& \frac{\Phi^2\delta_{ab}}{n + 4}\biggr\{\rho - 3p_e + 4p_i\biggl\}.
\end{eqnarray}

The components of the unimodular gravitational tensor, for a flat LFRW metric, are:
\begin{eqnarray}
E_{00} &=& \frac{1}{n + 4}\biggr\{ - 3(n + 2)\dot H - 3nH^2 - n(n + 2)\frac{\ddot\Phi}{\Phi} + 6nH\frac{\dot\Phi}{\Phi} 
+ n(n  - 1)\frac{\dot\Phi^2}{\Phi^2}\biggl\},\\
E_{ij} &=& \frac{a^2\delta_{ij}}{n + 4}\biggr\{ (n - 2)\dot H + 3nH^2 - 2n\frac{\ddot\Phi}{\Phi} 
+ n(n - 2)H\frac{\dot\Phi}{\Phi} - n(n  - 1)\frac{\dot\Phi^2}{\Phi^2}\biggl\},\\
E_{ab} &=& \frac{\Phi^2\delta_{ab}}{n + 4}\biggr\{ - 6(\dot H + 2H^2) - (n - 4)\biggr(\frac{\ddot\Phi}{\Phi} 
+ 3H\frac{\dot\Phi}{\Phi}\biggl) + 4(n  - 1)\frac{\dot\Phi^2}{\Phi^2}\biggl\}.
\end{eqnarray}

The equations of motion are:
\begin{eqnarray}
& &- 3(n + 2)\dot H - 3nH^2 - n(n + 2)\frac{\ddot\Phi}{\Phi} + 6nH\frac{\dot\Phi}{\Phi} + n(n  - 1)\frac{\dot\Phi^2}{\Phi^2} \nonumber\\
& & = 8\pi G\biggr\{(n + 3)\rho + 3 p_e + n p_i\biggl\};\\
& &(n - 2)\dot H + 3nH^2 - 2n\frac{\ddot\Phi}{\Phi} + n(n - 2)H\frac{\dot\Phi}{\Phi} - n(n  - 1)\frac{\dot\Phi^2}{\Phi^2}\nonumber \\ 
& &= 8\pi G\biggr\{\rho + (1 + n)p_e - n p_i\biggl\},\\
& & - 6(\dot H + 2H^2) - (n - 4)\biggr(\frac{\ddot\Phi}{\Phi} + 3H\frac{\dot\Phi}{\Phi}\biggl) + 4(n  - 1)\frac{\dot\Phi^2}{\Phi^2}\biggl\}\nonumber\\
& &= 8\pi G\biggr\{\rho - 3p_e + 4p_i\biggl\}.
\end{eqnarray}

\subsection{Solving the equations}

Now we will consider the five-dimensional KK theory ($n = 1$) and no pressure in the external and internal space.
The equations of motion under these hypothesis are:
\begin{eqnarray}
- 9\dot H - 3H^2 - 3\frac{\ddot\Phi}{\Phi} + 6H\frac{\dot\Phi}{\Phi}  &=& 32\pi G\rho;\\
-\dot H + 3H^2 - 2\frac{\ddot\Phi}{\Phi} - H\frac{\dot\Phi}{\Phi} &=& 8\pi G\rho;\\
- 6(\dot H + 2H^2) + 3\frac{\ddot\Phi}{\Phi} + 9H\frac{\dot\Phi}{\Phi}
&=& 8\pi G\rho.
\end{eqnarray}

These equation can be combine leading to,
\begin{eqnarray}
- 9\dot H - 3H^2 - 3\frac{\ddot\Phi}{\Phi} + 6H\frac{\dot\Phi}{\Phi}  &=& 24\pi G\rho,\\
\dot H + 3H^2 - \frac{\ddot\Phi}{\Phi} - 2H\frac{\dot\Phi}{\Phi}
&=& 0.
\end{eqnarray}
There are just two independent equations.
The system, even reduced to four dimensions, remains undetermined since we have just two equations for three variables:
$a$, $\Phi$ and $\rho$. 

One particular important case is when the internal space is static. In this case, we can solve the equations obtaining,
\begin{eqnarray}
a \propto t^{1/3}.
\end{eqnarray}
This is equivalent to standard solutions with stiff matter.

The system of equations is remains undetermined even if the pressure in the internal dimension is different from the pressure in 
the external dimension. For example, supposing $p_e = 0$ and $p_i = - \rho$, and considering again a static internal dimension, the scale factor behaves as
\begin{eqnarray}
a \propto t.
\end{eqnarray}
On the other hand, if again the internal dimension is static, but $p_e = - \rho$ and $p_i = 0$, we find a de Sitter phase in the external space:
\begin{eqnarray}
a \propto e^{Ht}, \quad H = \mbox{constant}.
\end{eqnarray}

One important general feature is that, after reduction to four dimension, the properties of the system of equations are not determined 
anymore by the combination $\rho + p$ as in the isotropic four dimensional unimodular theory, unless the pressure is the same 
in the internal and external spaces.

Now we compare the above results with the case of the theory in five-dimensions without the unimodular constraint. The equations, after reduction to four dimensions, are:
\begin{eqnarray}
G_{\mu\nu} - \biggr\{\frac{\Phi_{;\mu;\nu}}{\Phi} - g_{\mu\nu}\frac{\Box\Phi}{\Phi}\biggl\}  = 8\pi G \tilde T_{\mu\nu},\\
\frac{\Phi^2}{2}R\delta_{ab} = 8\pi G \tilde T_{ab}.
\end{eqnarray}
For a static internal dimension, the equations reduce to,
\begin{eqnarray}
3H^2 &=& 8\pi G \rho, \\
2\dot H + 3H^2 &=& - 8\pi Gp_e,\\
3\dot H + 6H^2 &=& 8\pi Gp_i.
\end{eqnarray}
One verifies directly that for the three cases above, with static internal dimension, the only consistent solution is the Minkowski spacetime, 
i.e. a static universe.

\section{Vacuum solution}

The equations.
\begin{eqnarray}
\label{fe1}
R_{\mu\nu} &=& 8\pi G T_{\mu\nu} + \frac{1}{\Phi}\biggr(\Phi_{;\mu;\nu} - g_{\mu\nu}\Box\Phi\biggl),\\
\label{fe2}
\Box\Phi &=& \frac{R}{3} - \frac{8\pi G}{3}T,\\
\label{fe3}
\frac{R_{;\nu}}{2}&=& 8\pi G\biggr({T^\mu_\nu}_{;\mu} + \frac{\Phi_{;\mu}}{\Phi}T^\mu_\nu\biggl).
\end{eqnarray}

FLRW metric,
\begin{eqnarray}
ds^2 = dt^2 - a(t)^2(dx^2 + dy^2 + dz^2).
\end{eqnarray}

Vacuum solutions.
\begin{eqnarray}
\label{vs1}
a(t) &=& a_0\cosh^{1/2} kt,\\
\label{vs2}
\phi(t) &=& \Phi_0\frac{\sinh kt}{\cosh^{1/2} kt}.
\end{eqnarray}
In these expressions, $k$ is positive integration constant.

The equation for the evolution of gravitational wave is:
\begin{eqnarray}
\ddot h_{ij} - \biggr(H - \frac{\dot\Phi}{\Phi}\biggl)\dot h_{ij} + \biggr\{\frac{q^2}{a^2} + 4H^2 - 2\biggr(\frac{\ddot\Phi}{\Phi} 
+ 3H\frac{\dot\Phi}{\phi}\biggl)\biggl\}h_{ij} = 0.
\end{eqnarray}
After inserting the solutions, and redefining $kt \rightarrow t$, the equation for gravitational wave reads,
\begin{eqnarray}
\label{gwe}
\ddot h_{ij} + \frac{\dot h_{ij}}{\cosh t\sinh t} + \biggr\{\frac{q^2}{\cosh t} - \frac{1}{\cosh^2 t} - 1\biggl\}h_{ij} = 0.
\end{eqnarray}
The equation (\ref{gwe}) display a divergent behaviour at the bounce, which occurs at $t = 0$.

\section{Solutions with matter}

In what concerns solutions with matter, the following point should be stressed at first. Equations (\ref{fe2}) and (\ref{fe3}) are obtained from
the single independent equation, namely (\ref{fe1}). In this case, we have only two independent equations of motion for $a$, $\phi$ and $\rho$. 
Hence, an extra condition is necessary. It seems that the most natural one is to impose the conservation of the $4$-dimensional energy-momentum tensor. 
However, an interesting alternative possibility consists in imposing the vanishing of the right-hand side of (\ref{fe3}):
\begin{eqnarray}
{T^\mu_\nu}_{;\mu} + \frac{\Phi_{;\mu}}{\Phi}T^\mu_\nu = 0.
\end{eqnarray}
With this condition imposed, the Ricci scalar is constant, as in the vacuum case, which results in a non-singular bouncing solution. 
 
This choice corresponds also to the following modification of the energy-momentum tensor:
\begin{eqnarray}
T^{\mu\nu} \rightarrow \frac{T^{\mu\nu}}{\Phi},
\end{eqnarray}
bringing (\ref{fe1}) to a form closer to the Brans-Dicke theory. It also may be viewed upon as a reminscence of the five-dimensional origin of the equations 
since the fifth dimension participates in the conservation of the matter/energy content.

Making this choice to complement the number of independent equations, we have the following set of equations:
\begin{eqnarray}
\label{em1}
- 3(\dot H + H^2) &=& 8\pi G \rho - 3H\frac{\dot \Phi}{\Phi}, \\
\label{em2}
\dot H + 3H^2 &=& 8\pi G p + \frac{\ddot\Phi}{\Phi} + 2H\frac{\dot\Phi}{\Phi},\\
\label{em3}
\dot\rho + 3H(\rho + p) + \frac{\dot \Phi}{\Phi}\rho &=& 0.
\end{eqnarray}
A simple linear relation between pressure and density is considered, since it covers some of the most important physical situation,
\begin{eqnarray}
p = \omega\rho, \quad \omega = \mbox{constant}.
\end{eqnarray}
The equation (\ref{em3}) can be easily solved,
leading to 
\begin{eqnarray}
\rho = \rho_0 a^{-3(1 + \omega)}\Phi, \quad \rho_0 = \mbox{constant}.
\end{eqnarray}

Equations (\ref{em1}) and (\ref{em2}) become,
\begin{eqnarray}
\label{em1A}
- 3(\dot H + H^2) &=& 8\pi G \frac{\rho_0}{a^{3(1 + \omega)}\Phi} - 3H\frac{\dot \Phi}{\Phi}, \\
\label{em2A}
\dot H + 3H^2 &=& 8\pi G\omega\frac{\rho_0}{a^{3(1 + \omega)}\Phi}  + \frac{\ddot\Phi}{\Phi} + 2H\frac{\dot\Phi}{\Phi}.
\end{eqnarray}
Both equations can be written as,
\begin{eqnarray}
\label{em3A}
- 3(\dot H + H^2)\Phi &=& 8\pi G \frac{\rho_0}{a^{3(1 + \omega)}} - 3H\dot \Phi, \\
\label{em4}
(\dot H + 3H^2)\Phi &=& 8\pi G\omega\frac{\rho_0}{a^{3(1 + \omega)}}  + \ddot\Phi + 2H\dot\Phi.
\end{eqnarray}
An inspection show that the solution for $\Phi$ may be written a,
\begin{eqnarray}
\Phi(t) = f(t) + \Phi_v(t),
\end{eqnarray}
where $\phi_v(t)$ is the vacuum solution (\ref{vs2}), and $f(t)$ is the function to be determined. Remark also that the scale factor 
in presence of matter is still given by (\ref{vs1}). Hence, after having determined $f(t)$ the consistency of equation (\ref{em3}) 
and (\ref{em4}) must be verified again.

\subsection{Solution for pressureless matter}

An explicit solution can be found for a pressureless matter, $p = 0$. 
Using the vacuum solution, the complete expression for the scalar field becomes,
\begin{eqnarray}
\phi = - A\cosh^{1/2}\tau + B\frac{\sinh \tau}{\cosh^{1/2}\tau}.
\end{eqnarray}
In this expression, $\tau = 2\sqrt{\Lambda}t$, $\Lambda$ and $B$ are arbitrary integration constants and
\begin{eqnarray}
A = - \frac{4\pi G\rho_0}{3\Lambda a_0^4}.
\end{eqnarray}
Remember that the scale factor is still given by
\begin{eqnarray}
a(\tau) = a_0\cosh^{1/2}\tau.
\label{bs1}
\end{eqnarray}
When matter is absent, $A = 0$ and the vacuum solution easy to be found.

 One characteristic of this pressureless matter solution is that the scalar field evolves again, as in the vacuum case, 
from negative to positive values passing by zero. However, the transition time does not coincide anymore with the moment 
when the scale factor bounces from contraction to expansion. The condition for a transition from negative values to positive values 
for $\phi$ is given by
\begin{eqnarray}
\frac{A}{B} = \tanh\tau_0,
\label{bs2}
\end{eqnarray}
$\tau_0$ being the transition time. Since $A > 0$ if matter is positive, the transition occurs for $\tau < 0 $ (contraction phase) 
if $B < 0$ but this implies that $\Phi$ remains negative during all evolution of the universe. If $\tau_0 > 0$, $\Phi$ transits 
from negative for positive values. If $\tau_0 = 0$, $A = 0$ and the vacuum solutions are recovered. Remark also that, 
in order to. have a transition from negative to positive value, $B$ must be positive and $B > A$.

Is this solution stable, at least with respect to tensorial perturbations? The evolution of tensorial modes, for presureless matter, 
is still governed by (\ref{gwe}) but with the background functions given by (\ref{bs1},\ref{bs2}). Choosing $B > 0$ and $B > A$, 
there is now a transition from negative to positive values that does not coincide anymore with the bouncing that occurs still at $\tau = 0$. 
This transition from negative to positive values of $\Phi$ is the dangerous moment for the stability of the system. 
The equation for the evolution of the tensorial modes at the vicinity of this transition time takes the form,
\begin{eqnarray}
\label{ve}
\ddot h_{ij} + a\frac{\dot h_{ij}}{x} + b\frac{h_{ij}}{x} = 0,
\end{eqnarray}
with,
\begin{eqnarray}
a &=& \coth^2\tau_0,\\
b &=& \tanh\tau_0\biggr(2 - 3\cosh^2\tau_0\biggl).
\end{eqnarray}
To obtain (\ref{ve}), une expansion around $\tau = \tau_0 + x$, $x << \tau_0$ has been made.
Remark that $b < 0$ since $\tau_0 > 0$ otherwise $\phi$ remains always negative.

The solution for (\ref{ve}) is given by,
\begin{eqnarray}
\label{ps}
h_{ij} = \epsilon_{ij} x ^{q/2}\biggr\{c_1 I_{|q|}(2\sqrt{|b|x}) + c_2 K_{|q|}(2\sqrt{|b|x})\biggl\},
\end{eqnarray}
with 
\begin{eqnarray}
q = - \frac{1}{\sinh^2\tau_0}.
\end{eqnarray}
The solution (\ref{ps}) diverges as $x \rightarrow 0$  due to the asymptotic behavior of the function $K_\nu(z)$. 
Even if the transition of $\Phi$ from negative to positive values does not occur at the bounce, the instability remains due to
flip of sign of $\Phi$.

The instability described above may change if $\phi$ remains always negative ($B < A$), with no change of sign 
during all evolution of the universe. In this case, it is not excluded that the instability reappears in the scalar sector 
(density perturbations), however a first inspection indicates that even in the scalar sector the configuration may be stable. 

A full perturbative analysis, including also the scalar sector and the consequences for structure formation, 
will be addressed in a separate study.

\section{Conclusions}
 
The main goal of the present article was to set up the general structure of the unimodular Kaluza-Klein gravity 
and its main features after the dimensional reduction to four dimensions.

As in the more familiar four-dimensional unimodular gravity, the unimodular constraint on the determinant of the five-dimensional metric 
restricts the invariance of the theory to the {\it transverse diffeomorphisms}. This leads to the possibility of a generalized conservation law 
for the energy-momentum tensor. If, however, the conservation 
of the five-dimensional energy momentum tensor is imposed, the usual KK five-dimensional equations are recovered with a cosmological constant 
in five dimensions. If conservation of the five dimensional energy-momentum tensor is not imposed, the KK structure can be recovered, 
but with an extra dynamical cosmological term, similarly 
to what happens in four dimensions.

Besides setting the general equations, some cosmological solutions in five-dimensions have been obtained.
In vacuum, the unimodular gravity becomes completely equivalent to GR in presence of a cosmological constant. Hence,
the Kasner-type solution of Ref. \cite{CD}, showing spontaneous compactification, is recovered in the KK unimodular gravity. 
In presence of matter the same features of the five dimensional GR theory in presence of a cosmological constant are reproduced in the unimodular context. 
However, UG gravity allows a generalized set of conservation laws: relaxing the usual conservation laws, new class of solutions appear. 
In order to implement these solutions an ansatz must be imposed, what is equivalent to introduce a dynamical cosmological term. In particular, 
in the cosmological context, we have shown that solutions with a constant internal dimension are possible even in presence of matter, 
what is not allowed in the usual GR context. 

Another interesting feature is the coupling of the gravity sector to the matter content when the five-dimensional unimodular gravity 
is reduced to four dimension. 
As was shown, for example, in Ref. \cite{velten}, in four dimensional GR unimodular gravity, the equations describing the evolution of universe, 
in the absence of usual conservation laws, depend only on the combination $\rho + p$ which may be identified, from a thermodynamical point of view,
with the enthalpy function. 
This remains true in the KK unimodular gravity in a five-dimensional isotropic and homogenous universe (which is not realistic, evidently). 
However, after reducing to four dimensions (or equivalently, by considering a Kasner-type structure in five dimensions), this identification 
is not verified any more, and the pressure acquires a specific role, as in the usual GR context.

In the present analysis we have explored cosmological solutions in absence of the electromagnetic field coming from the KK structure, 
which is required to have isotropy in the three spatial dimensional of the external space. For the KK unimodular gravity the reduction to four-dimension 
preserves the same coupling to the electromagnetic field emerging from the five-dimensional metric. This is related to the traceless character
 of the electromagnetic field. The coupling of the moduli field $\Phi$ associated to the fifth dimension, on the other hand, may possible bring new structures. 
One example, besides the cosmological scenario described here, may be the static, spherically symmetric configurations. Black holes may appear, 
for example, in the Brans-Dicke theory and in the usual KK theory with and without an electromagnetic field \cite{k1,k2,clement}, but only in the phantom regime. 
This general feature may change in the KK unimodular gravity. This problem deserves to be studied. For the cosmological solution found here, 
we may also expect peculiar features at perturbative level. Finally, the effective theory in four dimension, and the its coupling to matter emerging 
from KK unimodular, must be further developed since it corresponds to a new implementation of the coupling of the scalar field and gravity after reduction 
to the usual four-dimensional space-time.

\vskip 0.4cm
\indent
\hskip 1cm
{\bf APPENDIX I}
\vskip 0.3cm
\indent
Let us derive the set of general formulas for metrics, connections and curvature in $5$ dimensions, 
with all the $15$ degrees of freedom present. The calculus in coordinates turns out to be quite complicated, 
but introducing the non-holonomic local frames simplifies the computations considerably.

In the non-holonomic local frame the basis of $1$-forms is chosen to be:

\begin{equation}
\theta^{\mu} = d x^{\mu}, \; \; \; \; \theta^5 = dx^5 + k \; A_{\mu} dx^{\mu},
\label{thetas}
\end{equation}
(Note that the constant $k$ must have the dimension of length ($cm$) in order to ensure the uniformity with space variables $x^{\mu}$.  

The dual vector fields, satisfying $\theta^A ({\cal{D}}_B) = \delta^A_B$:
\begin{equation}
{\cal{D}}_{\mu} = \partial_{\mu} - k \; A_{\mu} \partial_5, \; \; \; \; {\cal{D}}_5 = \partial_5.
\label{CalD}
\end{equation}
Introducing transition matrices $U^A_B$ and ${\overset{-1}{U^B_C}}$ such that $\theta^A = U^A_B dx^B, \; {\cal{D}}_C = {\overset{-1}{U^D_C}} \partial_D $
we can write:
$$U^{\mu}_{\nu} = \delta^{\mu}_{\nu}, \; \; \; U^{\mu}_5 = 0, \; \; U^5_{\mu} = k A_{\mu}, \; \; U^5_5 = 1;$$
\begin{equation}
{\overset{-1}{U^{\mu}_{\nu}}} = \delta^{\mu}_{\nu}, \; \; {\overset{-1}{U^{5}_{\nu}}} = - k A_{\nu}, \; \; {\overset{-1}{U^{\mu}_{5}}} = 0 
\; \;  {\overset{-1}{U^{5}_{5}}} = 1.
\label{UUinv}  
\end{equation}

The metric tensor expressed in the non-holonomic frame is easily deduced from the square of the  $5$-dimensional length element,
taking on the following form:
\begin{equation}
ds^2 = g_{\mu \nu} dx^{\mu} dx^{\nu} - \Phi^2 \left[ dx^5 + k \; A_{\mu} dx^{\mu} \right] \left[ dx^5 + k \; A_{\nu} dx^{\nu} \right]
\label{deessquare}
\end{equation}
leading to the following $5 \times 5$ matrix representation:
\begin{equation}
g^{AB} = \begin{pmatrix} g_{\mu \nu} + k^2 \Phi^2 A_{\mu} A_{\nu} & - k \Phi^2 A_{\nu} \cr - k \Phi^2 A_{\mu} & - \Phi^2 \end{pmatrix} 
\label{gABfive}
\end{equation}
The inverse matrix becomes then:
\begin{equation}
g^{BC} = \begin{pmatrix} g^{\nu \lambda}  &  k A_{\lambda} \cr k  A_{\nu} & - \Phi^{-2} + k^2 A^{\nu} A^{\lambda} \end{pmatrix} 
\label{invgABfive}
\end{equation}
One easily checks that $g_{AB} g^{BC} = \delta^A_C.$
The simplest and most elegant way to evaluate the connection coefficients and the components of the Riemann tensor   
is to use the non-holonomic frame $\theta^A$ and its dual basis of derivations (vector fields) ${\cal{D}}_B, \; \; A, B = 1,2...5.$

We need to know the commutators of non-holonomic derivations. We have:
\begin{equation}
\left[ {\cal{D}}_A, {\cal{D}}_B \right] = C_{AB}^E \; {\cal{D}}_E,
\label{commDADB}
\end{equation}
where
\begin{equation} 
C^{5}_{\mu \nu} = C_{\mu \nu 5} = - k \; F_{\mu \nu} = -k \; (\partial_{\mu} A_{\nu} - \partial_{\nu} A_{\mu} ).
\end{equation}
%

We have then the connection coefficients in the non-holonomic basis:
\begin{equation}
{\hat{\Gamma}}^{C}_{AB} = \frac{1}{2} {\hat{g}}^{CE} \left[ {\cal{D}}_A g_{BE} + {\cal{D}}_B g_{AE} - {\cal{D}}_E g_{AB} \right] + 
{\hat{g}}^{CE} \left[ C_{EAB} + C_{EBA} - C_{BAE} \right]
\end{equation}
where the ``hat'' refers to the components with respect to the anholonomic frame.

The only non vanishing connection coefficients are then the following:
\begin{equation}
{\hat{\Gamma}}^{\mu}_{\nu \lambda} = \Gamma^{\mu}_{\nu \lambda}, \; \; {\hat{\Gamma}}^{\mu}_{\nu 5} = {\hat{\Gamma}}^{\mu}_{5 \nu} = - \frac{1}{2} k F^{\mu}_{\; \; \nu},
\; \;  {\hat{\Gamma}}^{5}_{\nu \lambda} = - {\hat{\Gamma}}^{5}_{\lambda \nu} = \frac{1}{2} k F_{\lambda \nu},
\label{Gammahat}
\end{equation}

The five-dimensional Riemann tensor expressed in a non-holonomic frame is:

\begin{equation}
{\hat{R}}_{AB \; \; \; D}^C = {\cal{D}}_A {\hat{\Gamma}}^C_{BD} - {\cal{D}}_B {\hat{\Gamma}}^C_{AD} + {\hat{\Gamma}}^C_{AF} {\hat{\Gamma}}^F_{BD}
- {\hat{\Gamma}}^C_{BF} {\hat{\Gamma}}^F_{AD} - C^F_{AB} {\hat{\Gamma}}^C_{FD} 
\label{Riemanntensor}
\end{equation}
The Ricci tensor and the curvature scalar in $5$ dimensions are calculated as usual,
\begin{equation}
{\hat{R}}_AD = {\hat{R}}_{AC \; \; \; D}^C, \; \; \; \; \; {\hat{R}} = {\hat{g}}^{AB} {\hat{R}}_{AB}.
\label{RicciR}
\end{equation}

The resulting expression for the $5$-dimensional curvature is quite simple indeed:

\begin{equation}
{\hat{R}} = {\overset{4}{R}} - \frac{1}{4} \Phi^2 \; F_{\mu \nu} F^{\mu \nu} - \frac{2}{3 \Phi^2} g^{\mu \nu} \partial_{\mu} \Phi \partial_{\nu} \Phi.
\label{RexplA}
\end{equation}

\vskip 0.3cm
\indent
\hskip 1cm
{\bf APPENDIX II}
\indent

%

In $1980$ Chodos and Detwiler \cite{CD} proposed a vacuum Kasner-type cosmological solution in the $5$-dimensional Kaluza-Klein space.
The metric element for this model was
\begin{equation}
ds^2 =dt^2 - {\sqrt{t}} \left[ dx^2 + dy^2 + dz^2 \right] - \frac{1}{\sqrt{t}} \rho^2 d \chi^2.
\label{cosmods}
\end{equation}
where the last angular variable $\chi$ comes from the fifth cyclic dimension, and $\rho$ is its radius. 

This metric can be generalized to more extra dimensions ($n$ to be more precise):

\begin{equation}
ds^2 = dt^2 - {\displaystyle{\sum_{i=1}^3}} t^{2 k_i} (dx^i)^2 - {\displaystyle{\sum_{a=4}^{3+n}}} t^{2 k_a} (dy^a)^2
\label{Kasnercosm}
\end{equation}
satisfying the following conditions:
\begin{equation}
{\displaystyle{\sum_{i=1}^3}} k_i + {\displaystyle{\sum_{a=4}^{3+n}}} k_a = 1, 
\end{equation}
\begin{equation}
{\displaystyle{\sum_{i=1}^3}} k_i^2 + {\displaystyle{\sum_{b=4}^{3+n}}} k_b^2 = 1
\end{equation}

The Friedmann-Robertson-Walker metric can be naturally generalized if we assume that the extra space dimensions 
form a compact spherically symmetric manifold. Then the overall metric can be derived from the following line element squared:
\begin{equation}
ds^2 = dt^2 - R_d^2 (t) \; g_{ij } dx^i dx^j - R_n^2 (t) g_{ab} dy^a dy^b,
\label{FRWplus}
\end{equation}
with two time-dependent scale factors, $R_d (t)$ for the space dimensions of our space-time, $d=3$, and $R_n (t)$ for the  
internal $n$-dimensional compact space - most usually, a $n$-dimensional sphere.

This ansatz yields the following Ricci tensor:
$$R_{00} = - \left[ 3 \frac{{\ddot{R}}_d}{R_d} + n \frac{{\ddot{R}}_n}{R_n} \right],$$
$$ R_{ij} = \left[ \frac{2k_d}{R_d^2} + \frac{d}{dt} \left( \frac{{\dot{R}_d}}{R_d} \right) + \frac{{\dot{R_d}}}{R_d} 
\left(3 \frac{{\dot{R_d}}}{R_d} + n \frac{{\dot{R_n}}}{R_n}\right) \right] g_{ij}, $$
$$ R_{ab} = \left[ \frac{2k_n}{R_n^2} + \frac{d}{dt} \left( \frac{{\dot{R}_n}}{R_n} \right) + \frac{{\dot{R_n}}}{R_n} 
\left(3 \frac{{\dot{R_d}}}{R_d} + n \frac{{\dot{R_n}}}{R_n}\right) \right] g_{ab}, $$

In $1985$ D. Sahdev \cite{Sahdev} obtained solutions of this system with several perfect fluids added on the right-hand side. The nice feature
was that $R_d$ was increasing with time, and $R_n$ decreasing. However, instead of stabilizing at some small but finite value,
as physics would require, the internal radius $R_n$ tended to zero. 

\bigskip

\noindent
{\bf Acknowledgement:} JCF thanks CNPq and FAPES for partial financial support. {\textcolor{violet} 
RK would like to acknowledge Philip D. Mannheim's valuable comments 
and suggestions }

\end{document}